\pdfoutput=1
\RequirePackage{ifpdf}
\ifpdf 
\documentclass[pdftex]{sigma}
\else
\documentclass{sigma}
\fi

\numberwithin{equation}{section}

\begin{document}

\newcommand{\arXivNumber}{2010.15273}

\renewcommand{\PaperNumber}{004}

\FirstPageHeading

\ShortArticleName{Ladder Operators and Hidden Algebras for Shape Invariant.~I}

\ArticleName{Ladder Operators and Hidden Algebras\\ for Shape Invariant Nonseparable\\ and Nondiagonalizable Models
with Quadratic\\ Complex Interaction. I.~Two-Dimensional Model}

\Author{Ian MARQUETTE~$^{\rm a}$ and Christiane QUESNE~$^{\rm b}$}

\AuthorNameForHeading{I.~Marquette and C.~Quesne}

\Address{$^{\rm a)}$~School of Mathematics and Physics, The University of Queensland,\\
\hphantom{$^{\rm a)}$}~Brisbane, QLD 4072, Australia}
\EmailD{\href{mailto:i.marquette@uq.edu.au}{i.marquette@uq.edu.au}}

\Address{$^{\rm b)}$~Physique Nucl\'eaire Th\'eorique et Physique Math\'ematique,
Universit\'e Libre de Bruxelles,\\
\hphantom{$^{\rm b)}$}~Campus de la Plaine CP229, Boulevard~du Triomphe, B-1050 Brussels, Belgium}
\EmailD{\href{mailto:christiane.quesne@ulb.be}{christiane.quesne@ulb.be}}

\ArticleDates{Received September 01, 2021, in final form January 03, 2022; Published online January 14, 2022}

\Abstract{A shape invariant nonseparable and nondiagonalizable two-dimensional model with quadratic complex interaction, first studied by Cannata, Ioffe, and Nishnianidze, is re-examined with the purpose of exhibiting its hidden algebraic structure. The two operators~$A^+$ and~$A^-$, coming from the shape invariant supersymmetrical approach, where $A^+$ acts as a raising operator while~$A^-$ annihilates all wavefunctions, are completed by introducing a novel pair of operators $B^+$ and $B^-$, where $B^-$ acts as the missing lowering operator. These four operators then serve as building blocks for constructing ${\mathfrak{gl}}(2)$ generators, acting within the set of associated functions belonging to the Jordan block corresponding to a given energy eigenvalue. This analysis is extended to the set of Jordan blocks by constructing two pairs of bosonic operators, finally yielding an ${\mathfrak{sp}}(4)$ algebra, as well as an ${\mathfrak{osp}}(1/4)$ superalgebra. Hence, the hidden algebraic structure of the model is very similar to that known for the two-dimensional real harmonic oscillator.}


\Classification{81Q05; 81Q60; 81R12; 81R15}

\section{Introduction}

During the last twenty years, there has been much interest in non-Hermitian Hamiltonians that under definite assumptions have a real spectrum. Most of them enjoy unbroken PT-invariance \cite{bender05, bender07, bender98} or, more generally, are endowed with the so-called pseudo-Hermiticity property, i.e., are such that $\eta H \eta^{-1} = H^{\dagger}$ with $\eta$ a Hermitian invertible operator \cite{mosta02a, mosta10}. For such systems, a suitable description of Hilbert space is given in terms of a biorthogonal basis consisting of the eigenstates $\Psi_n$ and $\tilde{\Psi}_n$ of $H$ and $H^{\dagger}$, respectively. The concept of pseudo-Hermiticity was introduced a long time ago by Pauli as generalized Hermiticity \cite{pauli43} and later on by Scholtz, Geyer, and Hahne as quasi-Hermiticity \cite{scholt92}.

If most studies have been carried out for one-dimensional systems with complex potential, there have also been some works on two- and three-dimensional systems (see, e.g., \cite{bender01, fer16, ioffe, nana}). In the present series of papers, we plan to re-examine two of them dealing with a quadratic complex interaction in two \cite{cannata10} or three dimensions \cite{barda}.

Both have the property of being exactly solvable although they are not amenable to separation of variables for the choice of parameters that is made. Their exact solvability is due to the fact that they satisfy the property of shape invariance~\cite{genden}, which was initially developed in one-dimensional supersymmetric quantum mechanics with real potential \cite{bagchi, cooper, junker} and was later on generalized to higher-dimensional systems~\cite{cannata02}. A Hamiltonian $H(\mathbf{x};a)$, depending on some variables $\mathbf{x}$ and some parameter $a$, is said to be shape invariant if it satifies supersymmetric intertwining relations with some operators $Q^{\pm}$, $H(\mathbf{x};a) Q^+ = Q^+ [H(\mathbf{x};\tilde{a}) + R(a)]$ and $Q^- H(\mathbf{x};a) = [H(\mathbf{x};\tilde{a}) + R(a)] Q^-$, where $\tilde{a} = \tilde{a}(a)$ is some function of $a$ and $R(a)$ does not depend on $\mathbf{x}$. The models studied in \cite{barda, cannata10} realize the simplest form of shape invariance, wherein $\tilde{a} = a$ and $R(a)$ is a constant and which leads to an equidistant spectrum with the spacing equal to $R(a)$ (the so-called `oscillator-like' or `self-isospectral' shape invariance).

The models of \cite{barda, cannata10} have the additional property that their Hamiltonian is not diagonalizable, which means that their (exactly known) wavefunctions do not realize a resolution of identity or, in others words, do not form a complete basis. The excited-state wavefunctions are actually self-orthogonal and have to be accompanied by a set of associated functions completing the Jordan blocks. This results in an extended biorthogonal basis, as well known for some one-dimensional pseudo-Hermitian Hamiltonians \cite{mosta02b, mosta02c}.

The main purpose of this series of papers is to show that, although the nonseparable and nondiagonalizable models with quadratic complex interaction of \cite{barda, cannata10} look rather more complicated than the two- and three-dimensional real harmonic oscillator models~\cite{mosh}, they are endowed with the same kind of symmetries as the latter. To bring this important property to light, it will prove convenient to build some novel ladder operators, which will complete the operators already known from shape invariance.

In the present paper, we will deal with the two-dimensional model, for which a set of associated functions has been determined in detail in \cite{cannata10}. In Section~\ref{section2}, we review the main results obtained there. In Section~\ref{section3}, we construct a set of additional ladder operators and calculate their action on the associated functions. In Section~\ref{section4}, we introduce some operators acting within a Jordan block and, from them, we build a realization of the ${\mathfrak{gl}}(2)$ algebra. In Section~\ref{section5}, such a construction is extended to the whole set of Jordan blocks, thereby giving rise to realizations of the ${\mathfrak{sp}}(4)$ algebra and of the ${\mathfrak{osp}}(1/4)$ superalgebra. Section~\ref{section6} then contains the conclusion.

\section{Shape invariant model with quadratic complex interaction}\label{section2}

\subsection{Shape invariant nonseparable model and its exact solvability}\label{section2.1}

Let us consider the two-dimensional model with complex oscillator Hamiltonian
\begin{gather}
 H = - \partial_1^2 - \partial_2^2 + \omega_1^2 x_1^2 + \omega_2^2 x_2^2 + 2 {\rm i}g x_1 x_2,
 \label{eq:H}
\end{gather}
where $\omega_1$, $\omega_2$, and $g$ are three real parameters. For generic values of the latter, the corresponding Schr\"odinger equation
\begin{gather*}
 H \Psi(\mathbf{x}) = E \Psi(\mathbf{x})
\end{gather*}
can be separated into two differential equations by performing a linear complex transformation of variables $x_1$, $x_2$~\cite{nana}.

As observed in \cite{cannata10}, this is not possible if the coupling constant is $g = \pm \big(\omega_1^2 - \omega_2^2\big)/2$ because the Jacobian of the transformation then vanishes. For $g = - \big(\omega_1^2 - \omega_2^2\big)/2$, for instance, Hamiltonian~(\ref{eq:H}) can be rewritten as
\begin{gather}
 H = - 4 \partial_z \partial_{\bar{z}} + \lambda^2 z \bar{z} + g \bar{z}^2 = - 4 \partial_z \partial_{\bar{z}}
 +4a^2 z \bar{z} + 8ab \bar{z}^2 \label{eq:H-bis}
\end{gather}
in terms of the complex variables $z = x_1 + {\rm i} x_2$, $\bar{z} = x_1 - {\rm i} x_2$, and the parameters $g$ and $\lambda = \sqrt{\big(\omega_1^2 + \omega_2^2\big)/2}$, or the parameters $a = \lambda/2$ and $b = g/(4\lambda)$. Such a Hamiltonian satisfies pseudo-Hermiticity with $\eta$ chosen as $P_2$, where $P_2$ is the operator changing $x_2$ into $-x_{2}$.

With the operators
\begin{gather*}
 A^{\pm} = \partial_z \mp a \bar{z},
\end{gather*}
Hamiltonian (\ref{eq:H-bis}) satisfies the properties
\begin{gather*}
 H A^+ = A^+ (H+4a), \qquad A^- H = (H+4a) A^-,
\end{gather*}
or
\begin{gather}
 [H, A^{\pm}] = \pm 4a A^{\pm}, \label{eq:H-A}
\end{gather}
characteristic of self-isospectral supersymmetry. Hence, it has an oscillator-like spectrum \cite{cannata10}
\begin{gather*}
 E_n = 4a (n+1), \qquad n=0, 1, 2, \dots,
\end{gather*}
and its ground-state wavefunction is annihilated by $A^-$, while its excited-state ones are obtained from the latter by successive applications of $A^+$. The results read \cite{cannata10}
\begin{gather}
 \Psi_{n,0}(z,\bar{z}) = c_{n,0} \bar{z}^n {\rm e}^{-az\bar{z} - b\bar{z}^2}, \qquad n=0, 1, 2, \dots,
 \label{eq:w-f}
\end{gather}
where $c_{n,0}$ was determined in \cite{cannata10} as\footnote{Here, we assume for simplicity's sake that $b>0$. Negative values of $b$ would be easily dealt with in the same way.}
\begin{gather*}
 c_{n,0} = \sqrt{\frac{2a}{\pi}} 4^n (ab)^{n/2}.
\end{gather*}

The operator $A^+$ acts as a raising operator, i.e.,
\begin{gather*}
 A^+ \Psi_{n,0} = - \frac{1}{2} \sqrt{\frac{a}{b}} \Psi_{n+1,0}, \qquad n=0, 1, 2, \dots,
\end{gather*}
but, in contrast with what happens for the real oscillator, $A^-$ is not a lowering operator, since it annihilates not only the ground-state wavefunction, but also all the excited-state ones,
\begin{gather*}
 A^- \Psi_{n,0} = 0, \qquad n=0, 1, 2, \dots,
\end{gather*}
and it also has the unusual property of commuting with $A^+$,
\begin{gather}
 [A^-, A^+] = 0. \label{eq:A-A}
\end{gather}

\subsection[Nondiagonalizability of the model and construction of an extended biorthogonal basis]{Nondiagonalizability of the model and construction\\ of an extended biorthogonal basis}\label{section2.2}

For self-consistency, non-Hermitian Hamiltonians such as (\ref{eq:H-bis}) need a suitable modification of the scalar product and resolution of identity \cite{bender05, bender07, mosta02a, mosta10}. A new scalar product can be defined as
\begin{gather}
 \langle\langle \Psi | \Phi \rangle\rangle = \int (B\Psi) \Phi \, {\rm d}{\bf x}, \label{eq:sp}
\end{gather}
where $B$ is an antilinear operator commuting with $H$. Here, one may take $B = P_2 T$. Then the pseudo-Hermitian Hamiltonian $H$ becomes Hermitian under the new scalar product (\ref{eq:sp}). Since the wavefunctions $\Psi_{n,0}(z,\bar{z})$ are simultaneously the eigenfunctions of $P_2 T$ with unique eigenvalue~$+1$, the new scalar product~(\ref{eq:sp}) becomes an integral over the product of functions~$\Psi \Phi$, instead of $\int \Psi^* \Phi\, {\rm d}{\bf x}$ as in quantum mechanics with real potentials. Other choices can be made as non-Hermitian operators may exhibit unitary and antiunitary symmetries~\cite{fer14}.

As it was shown in \cite{cannata10}, the norms of the basis states (\ref{eq:w-f}) are given by
\begin{gather*}
 \langle\langle \Psi_{n,0} | \Psi_{n,0} \rangle\rangle = \frac{\pi c_{n,0}^2}{2a} \delta_{n,0},
\end{gather*}
which means that only the ground-state wavefunction is normalizable. All excited-state wavefunctions are self-orthogonal, which signals that one deals with a nondiagonalizable Hamiltonian \cite{mosta02b, mosta02c}. As a consequence, some associated functions must be introduced to complete the basis and to get a resolution of identity.

In the present case, it was shown in \cite{cannata10} that to $\Psi_{n,0}(z,\bar{z})$, $n\ge1$, one has to add the functions $\Psi_{n,m}(z,\bar{z})$, $m=1, 2, \dots, n$, defined by
\begin{gather*}
 (H - E_n) \Psi_{n,m} = \Psi_{n, m-1}, \qquad m=1, 2, \dots, n,
\end{gather*}
and which assume the explicit form
\begin{gather}
 \Psi_{n,m}(z,\bar{z}) = c_n (2ab)^{n-m} {\rm e}^{-az\bar{z} - b\bar{z}^2} \nonumber\\
 \hphantom{\Psi_{n,m}(z,\bar{z}) =}{}\times \sum_{i=0}^{n-m} \alpha_i^{(n-m)}
 (2m-n+i+1)_{2n-2m-i} \bar{z}^i (az+b\bar{z})^{2m-n+i}. \label{eq:associated}
\end{gather}
Here $(a)_k = a(a+1)\cdots(a+k-1)$ is a Pochhammer symbol and
\begin{gather}
 \alpha^{(k)}_0 = (-2)^k, \qquad \alpha^{(k)}_k = 2^{2k}, \qquad \alpha^{(k)}_i = \frac{(-2)^i (k-i+1)_i
 \alpha^{(k)}_0}{i!}, \qquad 0<i<k, \nonumber \\
 c_n = \frac{c_{n,0}}{(8ab)^n n!}. \label{eq:associated-bis}
\end{gather}
Similarly, the partner eigenfunctions $\tilde{\Psi}_{n,0}(z,\bar{z})$, i.e., the eigenfunctions of $H^{\dagger}$, which are~needed to complete a biorthogonal basis, are accompanied by their associated functions $\tilde{\Psi}_{n,m}(z,\bar{z})$, $m=1, 2, \dots, n$, which can be taken as
\begin{gather*}
 \tilde{\Psi}_{n,n-m} = \Psi^*_{n,m}, \qquad m=0, 1, \dots, n.
\end{gather*}
The scalar product in the extended biorthogonal basis is then
\begin{align*}
 \langle \langle \Psi_{n,m} | \Psi_{n',m'} \rangle \rangle & = \langle \tilde{\Psi}_{n,m} | \Psi_{n',m'} \rangle
 = \int \Psi_{n,m} \Psi_{n',m'} d{\bf x} \nonumber \\
 & = \delta_{n,n'} \delta_{m,n-m'}, \qquad m=0, 1, \dots, n, \quad m' = 0, 1, \dots, n',
\end{align*}
with the corresponding decompositions
\begin{gather*}
 I = \sum_{n=0}^{\infty} \sum_{m=0}^n | \Psi_{n,m} \rangle \rangle \langle \langle \Psi_{n,n-m} |
\end{gather*}
and
\begin{gather*}
 H = \sum_{n=0}^{\infty} \sum_{m=0}^n E_n | \Psi_{n,m} \rangle \rangle \langle \langle \Psi_{n,n-m} |
 + \sum_{n=0}^{\infty} \sum_{m=0}^{n-1} | \Psi_{n,m} \rangle \rangle \langle \langle \Psi_{n,n-m-1} |,
\end{gather*}
showing that $H$ is block diagonal, each block having dimensionality $n+1$.

\section{Additional ladder operators}\label{section3}

The lack of a lowering operator resulting from shape invariance leads us to introduce another set of operators, defined by
\begin{gather*}
 B^{\pm} = \partial_{\bar{z}} \mp az \mp 2b\bar{z},
\end{gather*}
where $B^-$ can provide us with such an operator. Its action on the set of functions $\Psi_{n,m}(z,\bar{z})$, $n=0, 1, 2, \dots$, $m=0, 1, \dots n$, defined in (\ref{eq:associated}) and (\ref{eq:associated-bis}), can indeed be easily shown to be given by
\begin{gather}
 B^- \Psi_{n,m} = \begin{cases}
 4n \sqrt{ab} \Psi_{n-1,0} & \text{if $m=0$}, \\
 \displaystyle 4(n-m) \sqrt{ab} \Psi_{n-1,m} + \frac{1}{2}\sqrt{\frac{b}{a}} \Psi_{n-1,m-1} &
 \text{if $m=1, 2, \dots, n$}.
 \end{cases} \label{eq:action-1}
\end{gather}
Similarly, that of $B^+$ is obtained as
\begin{gather}
 B^+ \Psi_{n,m} = - 4(m+1)\sqrt{ab} \Psi_{n+1,m+1} - \frac{1}{2} \sqrt{\frac{b}{a}} \Psi_{n+1,m},
 \qquad m=0,1, \dots, n,\label{eq3.2}
\end{gather}
for any $n = 0, 1, 2, \dots$.

In comparison, the action of the operators $A^{\pm}$ on the same functions is given by
\begin{gather}
 A^- \Psi_{n,m} = \begin{cases}
 0 & \text{if $m=0$}, \\
 \displaystyle \frac{1}{2} \sqrt{\frac{a}{b}} \Psi_{n-1,m-1} & \text{if $m=1, 2, \dots, n$},
 \end{cases}\label{eq3.3}
\end{gather}
and
\begin{gather}
 A^+ \Psi_{n,m} = - \frac{1}{2} \sqrt{\frac{a}{b}} \Psi_{n+1,m}, \qquad m=0, 1, \dots, n,
 \label{eq:action-2}
\end{gather}
for any $n=0, 1, 2, \dots$.

These relations can be completed by the set of commutators
\begin{gather}
 [H, B^{\pm}] = \pm 4a B^{\pm} \pm 8b A^{\pm},\qquad
 [B^-, B^+] = - 4b,
 \label{eq:H-B}
\end{gather}
as well as
\begin{gather}
 [A^{\pm}, B^{\pm}] = 0, \qquad [A^{\pm}, B^{\mp}] = \pm 2a. \label{eq:A-B}
\end{gather}

As we plan to show in the next two sections, the four operators $A^{\pm}$ and $B^{\pm}$ will provide us with building blocks to bring the hidden algebraic structure of the model to light.

\section[Operators acting within a Jordan block and realization of gl(2)]{Operators acting within a Jordan block\\ and realization of $\boldsymbol{{\mathfrak{gl}}(2)}$}\label{section4}

Apart from $H$, one can form four operators that do not change the $n$ value and which therefore act within the corresponding Jordan block,
\begin{gather*}
 R = A^+ A^-, \qquad S = B^+ B^-, \qquad T = A^+ B^- - B^+ A^-, \qquad U = A^+ B^- + B^+ A^-.
\end{gather*}
In terms of the variables $z$ and $\bar{z}$, they can be written as
\begin{gather*}
\begin{split}
& R = \partial_z^2 - a^2 \bar{z}^2, \qquad S = \partial_{\bar{z}}^2 - (az+2b\bar{z})^2 + 2b, \\
& T = 2(az+2b\bar{z}) \partial_z - 2a \bar{z} \partial_{\bar{z}}, \qquad U = 2 \partial_z \partial_{\bar{z}}
 - 2a^2 z\bar{z} - 4ab \bar{z}^2 + 2a,
\end{split}
\end{gather*}
from which we note that
\begin{gather}
 U = - \frac{1}{2} H + 2a. \label{eq:U-H}
\end{gather}

From equations (\ref{eq:H-A}), (\ref{eq:A-A}), (\ref{eq:H-B}), (\ref{eq:A-B}), and (\ref{eq:U-H}), we directly obtain the following commutation relations
\begin{gather*}
 [H, R] = 0, \qquad [H, S] = 8bT, \qquad [H, T] = - 16bR, \qquad [H, U] = 0, \\
 [R, S] = - 2aT, \qquad [R, T] = 4aR, \qquad [R, U] = 0, \\
 [S, T] = - 4aS + 4bU, \qquad [S, U] = 4bT, \qquad [T, U] = - 8bR,
\end{gather*}
which show that the four operators $R$, $S$, $T$, and $U$ (or $H$) generate some Lie algebra. The latter is easily identified as ${\mathfrak{gl}}(2)$ because the combined operators
\begin{gather*}
 J_0 = \frac{1}{2}(E_{11} - E_{22}) = \frac{T}{4a}, \qquad J_+ = E_{12} = - \frac{1}{16ab} \left(S +
 \frac{b^2}{a^2} R - \frac{b}{a} U\right), \\
 J_- = E_{21} = - \frac{4b}{a} R, \qquad K = E_{11} + E_{22} = \frac{1}{2a} \left(\frac{2b}{a} R - U + 2a
 \right) = \frac{1}{4a} \left(H + \frac{4b}{a} R\right)
\end{gather*}
satisfy the commutation relations
\begin{gather*}
 [K, J_0] = [K, J_{\pm}] = 0, \qquad [J_0 , J_{\pm}] = \pm J_{\pm}, \qquad [J_+, J_-] = 2J_0,
\end{gather*}
or
\begin{gather*}
 [E_{ij}, E_{kl}] = \delta_{j,k} E_{il} - \delta_{i,l} E_{kj}.
\end{gather*}

At this stage, it is worth pointing out an important difference with respect to the symmetry algebra u(2) of the two-dimensional Hermitian harmonic oscillator. For the latter, the Hamiltonian is equal to the first-order Casimir operator~$K$ up to some multiplicative constant and it is therefore superintegrable. In the present case, $H$ turns out to be a linear combination of~$K$ and~$J_-$ and only $R \propto J_-$ commutes with it, showing that~$H$ is only integrable.

Furthermore, from equations~(\ref{eq:action-1}), (\ref{eq:action-2}), it is straightforward to show that
\begin{gather*}
 R \Psi_{n,m} = \begin{cases}
 0 & \text{if $m=0$}, \\
\displaystyle - \frac{a}{4b} \Psi_{n, m-1} & \text{if $m=1, 2, \dots, n$},
 \end{cases}
\\
 S \Psi_{n,m} = \begin{cases}
 - 2bn \Psi_{n,0} - 16abn \Psi_{n,1} & \text{if $m=0$}, \\
\displaystyle - \frac{b}{4a} \Psi_{n,m-1} - 2bn \Psi_{n,m} - 16ab(n-m)(m+1) \Psi_{n,m+1} & \text{if $m=1, 2, \dots, n$},
 \end{cases}
\\
 T \Psi_{n,m} = - 2a (n-2m) \Psi_{n,m},
\end{gather*}
and
\begin{gather*}
 U \Psi_{n,m} = \begin{cases}
 -2an \Psi_{n,0} & \text{if $m=0$}, \\
\displaystyle -2an \Psi_{n,m} - \frac{1}{2} \Psi_{n,m-1} & \text{if $m=1,2, \dots, n$},
 \end{cases}
\end{gather*}
from which we directly get
\begin{gather*}
\begin{split}
& J_0 \Psi_{n,m}= \left(m- \frac{n}{2}\right) \Psi_{n,m}, \\
& J_+ \Psi_{n,m}= (n-m)(m+1) \Psi_{n,m+1}, \\
& J_- \Psi_{n,m} = \begin{cases}
 0 & \text{if $m=0$}, \\
 \Psi_{n,m-1} & \text{if $m=1, 2, \dots, n$},
 \end{cases} \\
& K \Psi_{n,m}= (n+1) \Psi_{n,m}.
\end{split}
\end{gather*}

It is therefore clear that the renormalized functions
\begin{gather}
 \Phi_{j, \mu} = \sqrt{\frac{m!}{(n-m)!}} \Psi_{n,m}, \label{eq:Phi}
\end{gather}
where $j = \frac{n}{2}$ and $\mu = \frac{1}{2}(2m-n)$, running over $\big\{{-}\frac{n}{2}, -\frac{n}{2}+1, \dots, \frac{n}{2}\big\}$ or $\{-j, -j+1, \dots, j\}$, fulfil the usual relations characteristic of an irreducible representation~$j$ of ${\mathfrak{sl}}(2)$, namely
\begin{gather*}
 J_0 \Phi_{j,\mu} = \mu \Phi_{j,\mu}, \qquad J_{\pm} \Phi_{j,\mu} = \sqrt{(j\mp\mu)(j\pm\mu+1)} \Phi_{j,\mu\pm1},
\end{gather*}
the first-order ${\mathfrak{gl}}(2)$ Casimir operator satisfying the relation
\begin{gather*}
 K \Phi_{j,\mu} = (2j+1) \Phi_{j,\mu}.
\end{gather*}

\section[Operators acting on the whole set of Jordan blocks and realizations of sp(4) and osp(1/2)]{Operators acting on the whole set of Jordan blocks\\ and realizations of $\boldsymbol{{\mathfrak{sp}}(4)}$ and $\boldsymbol{{\mathfrak{osp}}(1/2)}$}\label{section5}

To mix functions belonging to different Jordan blocks, it is necessary to reintroduce the two sets of operators $A^{\pm}$ and $B^{\pm}$. It is actually appropriate to combine them in order to get two commuting sets of bosonic operators $a_i^{\pm}$, $i=1$, 2, satisfying the commutation relations
\begin{gather}
 [a_i^-, a_j^+] = \delta_{i,j}, \qquad [a_i^-, a_j^-] = [a_i^+, a_j^+] = 0. \label{eq:a-com}
\end{gather}
Such operators can be defined as
\begin{gather}
 a_1^+ = \frac{1}{4a\sqrt{ab}} (b A^+ - a B^+), \qquad a_1^- = 2 \sqrt{\frac{b}{a}} A^-,
 \label{eq:a-def-1}
\end{gather}
and
\begin{gather}
 a_2^+ = - 2\sqrt{\frac{b}{a}} A^+, \qquad a_2^- = - \frac{1}{4a\sqrt{ab}} (b A^- - a B^-).
 \label{eq:a-def-2}
\end{gather}

In addition to the commutation relations (\ref{eq:a-com}), we may also consider the anticommutators of the operators $a_i^{\pm}$, $i=1$, 2. It is straightforward to show that those of $a_i^-$ and $a_j^+$ give back the operators $J_0$, $J_+$, $J_-$, and $K$, or $E_{ij}$, introduced in Section~4,
\begin{gather*}
\{a_1^-, a_1^+\}= \frac{1}{2a}\left(\frac{2b}{a} R + T - U + 2a\right) = K + 2J_0, \\
\{a_1^-, a_2^+\}= - \frac{8b}{a} R = 2 J_-, \\
\{a_2^-, a_1^+\}= - \frac{1}{8ab}\left(\frac{b^2}{a^2} R + S - \frac{b}{a} U\right) = 2 J_+, \\
\{a_2^-, a_2^+\}= \frac{1}{2a}\left(\frac{2b}{a} R - T - U + 2a\right) = K - 2 J_0,
\end{gather*}
or, equivalently,
\begin{gather}
 E_{ij} = \frac{1}{2} \{a_i^+, a_j^-\} = a_i^+ a_j^- + \frac{1}{2} \delta_{i,j}. \label{eq:a-anticom-1}
\end{gather}

Furthermore, the anticommutators of $a_i^{\pm}$ and $a_j^{\pm}$ provide us with some new operators
\begin{gather}
 D^+_{ij} = \frac{1}{2} \{a_i^+, a_j^+\} = a_i^+ a_j^+, \qquad D^-_{ij} = \frac{1}{2} \{a_i^-, a_j^-\} =
 a_i^- a_j^-, \label{eq:a-anticom-2}
\end{gather}
which can be rewritten in terms of $A^{\pm}$ and $B^{\pm}$ as
\begin{gather*}
D^+_{11}= \frac{1}{16a^3b}\big[b^2 (A^+)^2 - 2ab A^+ B^+ + a^2 (B^+)^2\big], \\
D^+_{12}= - \frac{1}{2a^2} \big[b (A^+)^2 - a A^+ B^+\big], \\
D^+_{22}= \frac{4b}{a} (A^+)^2, \\
D^-_{11}= \frac{4b}{a} (A^-)^2, \\
D^-_{12}= - \frac{1}{2a^2} \big[b (A^-)^2 - a A^- B^-\big], \\
D^-_{22}= \frac{1}{16a^3b} \big[b^2 (A^-)^2 - 2ab A^- B^- + a^2 (B^-)^2\big].
\end{gather*}

From their definition in terms of the bosonic operators $a_i^{\pm}$, $i=1, 2$, it is clear that the operators $E_{ij}$, $D^+_{ij}$, and $D^-_{ij}$ generate an ${\mathfrak{sp}}(4)$ algebra \cite{mosh} and that, together with the former, make rise to an ${\mathfrak{osp}}(1/4)$ superalgebra \cite{frappat}. We indeed get the following set of commutators
\begin{gather*}
\begin{split}
& [E_{ij}, D^+_{kl}]= \delta_{j,k} D^+_{il} + \delta_{j,l} D^+_{ik}, \\
& [E_{ij}, D^-_{kl}]= - \delta_{i,k} D^-_{jl} - \delta_{i,l} D^-_{jk}, \\
& [D^-_{ij}, D^+_{kl}]= \delta_{i,k} E_{lj} + \delta_{i,l} E_{kj} + \delta_{j,k} E_{li} + \delta_{j,l} E_{ki}, \\
& [D^{\pm}_{ij}, D^{\pm}_{kl}]= 0, \\
& [a_i^-, E_{jk}]= \delta_{i,j} a_k^-, \\
& [a_i^+, E_{jk}]= - \delta_{i,k} a_j^+, \\
& [a_i^{\pm}, D^{\mp}_{jk}]= \mp \delta_{i,j} a_k^{\mp} \mp \delta_{i,k} a_j^{\mp}, \\
& [a_i^{\pm}, D^{\pm}_{jk}]= 0,
\end{split}
\end{gather*}
in addition to the set of anticommutators given in equations~(\ref{eq:a-anticom-1}) and (\ref{eq:a-anticom-2}). The explicit expressions of the ${\mathfrak{osp}}(1/4)$ generators in terms of $z$ and $\bar{z}$ are given in Appendix~\ref{appendixA}.

It remains to determine the action of the operators introduced in this section on the functions $\Psi_{n,m}(z,\bar{z})$. From equations~(\ref{eq:action-1}), \eqref{eq3.2}, \eqref{eq3.3}, (\ref{eq:action-2}), (\ref{eq:a-def-1}), and (\ref{eq:a-def-2}), we directly get the relations
\begin{gather*}
 a_1^+ \Psi_{n,m}= (m+1) \Psi_{n+1,m+1}, \\
 a_1^- \Psi_{n,m}= \begin{cases}
 0 & \text{if $m=0$}, \\
 \Psi_{n-1,m-1} & \text{if $m=1, 2, \dots, n$},
 \end{cases} \\
 a_2^+ \Psi_{n,m}= \Psi_{n+1,m}, \\
 a_2^- \Psi_{n,m}= (n-m) \Psi_{n-1,m},
\end{gather*}
which imply
\begin{gather*}
 D^+_{11} \Psi_{n,m}= (m+1)(m+2) \Psi_{n+2,m+2}, \\
 D^+_{12} \Psi_{n,m}= (m+1) \Psi_{n+2,m+1}, \\
 D^+_{22} \Psi_{n,m}= \Psi_{n+2,m},
\end{gather*}
and{\allowdisplaybreaks
\begin{gather*}
 D^-_{11} \Psi_{n,m}= \begin{cases}
 0 & \text{if $m=0,1$}, \\
 \Psi_{n-2,m-2} & \text{if $m=2,3, \dots, n$},
 \end{cases} \\
 D^-_{12} \Psi_{n,m}= \begin{cases}
 0 & \text{if $m=0$}, \\
 (n-m) \Psi_{n-2,m-1} & \text{if $m=1, 2, \dots, n$},
 \end{cases} \\
 D^-_{22} \Psi_{n,m}= (n-m)(n-m-1) \Psi_{n-2,m}.
\end{gather*}

}

Finally, if instead of the set of $\Psi_{n,m}$ functions, we use the renormalized functions $\Phi_{j,\mu}$, defined in (\ref{eq:Phi}), we get the same relations as for the conventional two-dimensional model with real harmonic interaction, namely
\begin{gather*}
a_1^{\pm} \Phi_{j,\mu}= \sqrt{j+\mu+\frac{1}{2} \pm \frac{1}{2}} \Phi_{j\pm\frac{1}{2}, \mu\pm
 \frac{1}{2}}, \\
 a_2^{\pm} \Phi_{j,\mu}= \sqrt{j-\mu+\frac{1}{2} \pm \frac{1}{2}} \Phi_{j\pm\frac{1}{2}, \mu\mp
 \frac{1}{2}}, \\
 D^{\pm}_{11} \Phi_{j,\mu}= \sqrt{(j+\mu\pm1) (j+\mu+1\pm1)} \Phi_{j\pm1, \mu\pm1}, \\
 D^{\pm}_{12} \Phi_{j,\mu}= \sqrt{\left(j-\mu+\frac{1}{2}\pm\frac{1}{2}\right) \left(j+\mu+
 \frac{1}{2}\pm\frac{1}{2}\right)} \Phi_{j\pm1, \mu}, \\
 D^{\pm}_{22} \Phi_{j,\mu}= \sqrt{(j-\mu\pm1) (j-\mu+1\pm1)} \Phi_{j\pm1, \mu\mp1}.
\end{gather*}

\section{Conclusion}\label{section6}

In the present paper, we have re-examined the shape invariant nonseparable and nondiagonalizable two-dimensional model with quadratic complex interaction that was first studied in \cite{cannata10} with the purpose of exhibiting its hidden algebraic structure. In contrast with the usual Hermitian 2D harmonic oscillator, this hidden algebra is related with integrability and not superintegrability.

For such a purpose, we have first made up for the lack of lowering operator coming from shape invariance by introducing a new operator $B^-$, possessing such a property with respect to the whole set of associated functions. Together with its accompanying operator~$B^+$, it has provided us with a pair of operators, completing the couple of operators $A^+$ and $A^-$ coming from the shape invariant supersymmetric approach of~\cite{cannata10}.

By combining these four operators, we have then built the generators $E_{ij}$, $i,j=1,2$, of a~${\mathfrak{gl}}(2)$ algebra and shown that the set of associated functions $\{\Psi_{n,m} | m=0, 1, \dots, n\}$, belonging to the Jordan block corresponding to an energy eigenvalue $E_n = 4a (n+1)$, are basis functions for the irreducible representation $j=n/2$ of the corresponding ${\mathfrak{sl}}(2)$ subalgebra.

Such a construction has been enlarged to the set of Jordan blocks by introducing two pairs of bosonic operators $a_i^{\pm}$, $i=1,2$, obtained by linearly combining the operators $A^{\pm}$ and $B^{\pm}$. These bosonic operators have then served as building blocks for the operators $D^{\pm}_{ij}$, $i,j=1,2$, that generate an ${\mathfrak{sp}}(4)$ algebra together with the ${\mathfrak{gl}}(2)$ generators previously obtained. The whole set of operators $E_{ij}$, $D^{\pm}_{ij}$, $a_i^{\pm}$, $i,j=1,2$, has finally provided us with an ${\mathfrak{osp}}(1/4)$ superalgebra.

In conclusion, we have established that the model of \cite{cannata10} has a hidden algebraic structure very similar to that known for the two-dimensional real harmonic oscillator. In the next paper, we plan to extend such an analysis to the three-dimensional model of \cite{barda}.

\appendix\allowdisplaybreaks

\section[The osp(1/4) generators in terms of z and bar z]{The $\boldsymbol{{\mathfrak{osp}}(1/4)}$ generators in terms of $\boldsymbol{z}$ and $\boldsymbol{\bar{z}}$}\label{appendixA}

In this appendix, we present the explicit expressions of the ${\mathfrak{osp}}(1/4)$ generators, defined in Sections~\ref{section4} and~\ref{section5}, in terms of the variables $z$ and $\bar{z}$:
\begin{gather*}
 a_1^+ = \frac{1}{4a\sqrt{ab}} [b \partial_z - a \partial_{\bar{z}} + a(az+b\bar{z})], \\
 a_1^- = 2 \sqrt{\frac{b}{a}} (\partial_z + a\bar{z}), \\
 a_2^+ = - 2\sqrt{\frac{b}{a}} (\partial_z - a \bar{z}), \\
 a_2^- = - \frac{1}{4a\sqrt{ab}} [b \partial_z - a \partial_{\bar{z}} - a(az+b\bar{z})], \\
 J_0 = \frac{1}{2a} [(az+2b\bar{z}) \partial_z - a\bar{z}\partial_{\bar z}], \\
 J_+ = - \frac{1}{16a^3b} \big[(b\partial_z-a\partial_{\bar z})^2 - a^2 (az+b\bar{z})^2\big], \\
 J_- = - \frac{4b}{a} \big(\partial_z^2 - a^2 \bar{z}^2\big), \\
 K = \frac{1}{a^2} \big[b\partial_z^2 - a\partial_z\partial_{\bar z} + a^2\bar{z}(az+b\bar{z})\big], \\
 D^+_{11} = \frac{1}{16a^3b} \big[(b\partial_z-a\partial_{\bar z})^2 + 2a (az+b\bar{z})
 (b\partial_z-a\partial_{\bar z}) + a^2 (az+b\bar{z})^2\big], \\
 D^+_{12} = - \frac{1}{2a^2} \big[b\partial_z^2 - a\partial_z\partial_{\bar z} + a^2(z\partial_z+\bar{z}
 \partial_{\bar z}) - a^2\bar{z}(az+b\bar{z}) + a^2\big], \\
 D^+_{22} = \frac{4b}{a} \big(\partial_z^2 - 2a\bar{z}\partial_z + a^2\bar{z}^2\big), \\
 D^-_{11} = \frac{4b}{a} \big(\partial_z^2 + 2a\bar{z}\partial_z + a^2\bar{z}^2\big), \\
 D^-_{12} = - \frac{1}{2a^2} \big[b\partial_z^2 - a\partial_z\partial_{\bar z} - a^2(z\partial_z+\bar{z}
 \partial_{\bar z}) - a^2\bar{z}(az+b\bar{z}) - a^2\big], \\
 D^-_{22} = \frac{1}{16a^3b} \big[(b\partial_z-a\partial_{\bar z})^2 - 2a (az+b\bar{z})
 (b\partial_z-a\partial_{\bar z}) + a^2 (az+b\bar{z})^2\big].
\end{gather*}

\subsection*{Acknowledgments}

I.~Marquette was supported by Australian Research Council Future Fellowhip FT180100099. C.~Quesne was supported by the Fonds de la Recherche Scientifique - FNRS under Grant Number 4.45.10.08.

\pdfbookmark[1]{References}{ref}
\LastPageEnding

\end{document}